\documentclass{article}
\usepackage{hiph-art}
\volnumber{19} \issuenumber{1} \edyear{2004}                             
\frompage{000} \topage{000}                                              
\recrevdate{15 September 2005}                                              
\def\rightmark{Heavy Quarks and Heavy Quarkonia}
\def\leftmark{J.L. Nagle for the PHENIX Collaboration}
 
\title{Heavy Quarks and Heavy Quarkonia as Tests of Thermalization} 
\authors{ 
{J.L.Nagle$^1$ for the PHENIX Collaboration %
\index{Nagle, J.L.} 
}\\[2.812mm]
{\normalsize
\hspace*{-8pt}$^1$ University of Colorado,\\ 
Boulder, CO 80305, USA\\[0.2ex] 
}}
 
\abstract{We present here a brief summary of the presentation given at the 
``Quark Gluon Plasma Thermalization'' Workshop in Vienna, Austria in August 2005, 
directly following the International Quark Matter Conference in Hungary.  }

 
\makeindex
\begin{document}
 
\maketitle

\section{Introduction}\label{intro}
 
In the PHENIX White Paper~\cite{phenix_whitepaper}, we reported the following conclusions:  (1) At RHIC we have created
bulk matter at energy densities well above that predicted by lattice QCD for the transition to a Quark Gluon Plasma (QGP).  
(2)  The energy density is dominantly equilibrated at very early times ($< 2$ fm/c), which is when the 
energy density is highest.  (3) The bulk matter behaves collectively and as such has been described as a nearly
perfect fluid.  We want to push these conclusions further utilizing new data from the large statistics 
$Au+Au$ and $Cu+Cu$ running at RHIC, in particular on heavy quark dynamics and heavy quarkonia suppression or
lack thereof. 

Almost three years ago, some of us suggested that the PHENIX data on non-photonic electrons (presumably from heavy flavor
meson decay) may be consistent with charm thermalization and hydrodynamic flow~\cite{batsouli}.  At the time, many 
dismissed this hypothesis, and yet now this is the commonly held belief in the field and supported by new experimental
data.  The large charm quark mass means that only very strong interactions with high frequency can bring them into
equilibrium with the light quarks and gluons in the medium.  In a calculation by Teaney and Moore~\cite{teaney}, they calculate
the expected transverse momentum modifications ($R_{AA}$) and momentum anisotropy ($v_{2}$) for different charm
quark diffusion coefficients - put in as a free parameter in their calculation.  
Their calculations show that the suppression of high transverse momentum charm goes hand in
hand with an increase in the momentum anisotropy.  

Lattice QCD results show that the confining potential between heavy quarks is modified - screened - at high temperatures.
At sufficiently high temperatures, this screening should suppress bound state formation, such as the $J/\psi$.  However,
recent lattice results indicate that the $J/\psi$ spectral function show only modest modification near the critical
temperature, and thus may not be suppressed until significantly higher temperatures.

\section{Experimental Results}

The PHENIX Experiment was designed to measure electrons, muons, photons and hadrons utilizing rare
event triggers and high data acquisition throughput~\cite{nim_phenix}.  For the heavy quark and quarkonia results
detailed in this proceedings, we utilize the particle identification of electrons in two central spectrometers.  The
acceptance is around mid-rapidity $-0.35 < \eta < +0.35$ and electron-pion separation is achieved using a
Ring Imaging Cerenkov Counter and track matching to an Electromagnetic Calorimeter.  Also crucial for our measurement
is the very low radiation length ($< 0.4$\%), which keeps the photon conversion background low.  In addition, we identify
muons at forward rapidities $1.2 < |y| < 2.2$ in the PHENIX muon spectrometers through a series of interleaved
absorbers and active detectors.  

\subsection{Open Charm Results}

PHENIX has published results on open charm indicating that the total charm yield scales with the number
of binary collisions~\cite{charm_scaling}.  This indicates that charm production may be a ``hard process'' and not suffer 
large modification due to coherence effects.  Note that this result does not comment on modification of the distribution
of charmed hadrons, but only on the scaling of the integrated $dN/dy (0.5 < p_{T} < 4.0$~GeV) near mid-rapidity.  In
addition, PHENIX has published the first results at RHIC from a modest $Au+Au$  data sample from Run-2 
revealing a non-zero momentum anisotropy ($v_{2}$) for non-photonic electrons~\cite{charm_flow1}.

\begin{figure}[htb]
\vspace*{+0.0cm}
\insertplot{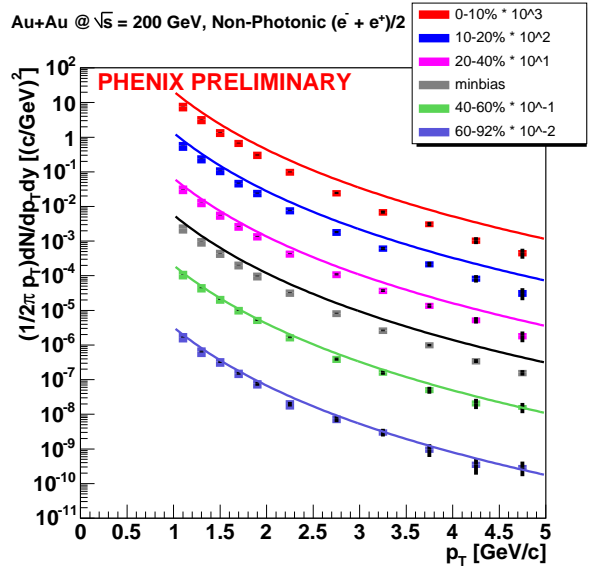}
\vspace*{-0.7cm}
\caption[]{PHENIX Preliminary $Au+Au$ $\sqrt{s_{NN}}$=200 GeV data for the invariant yield of non-photonic electrons versus transverse momentum
for various centrality selections.  The curve is a best fit to the proton-proton yield scaled up by the expected
number of binary collisions.}
\label{fig_elec_spectra}
\end{figure}

\begin{figure}[htb]
\vspace*{+0.0cm}
\insertplot{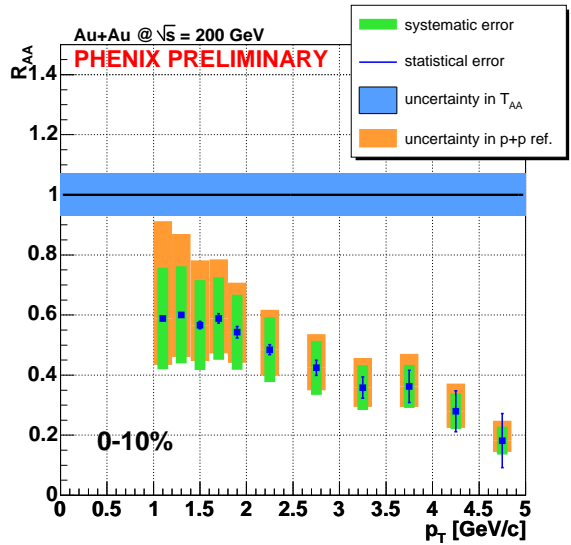}
\vspace*{-0.7cm}
\caption[]{PHENIX Preliminary central 0-10\% $Au+Au$ $\sqrt{s_{NN}}$=200 GeV suppression factor $R_{AA}$ for non-photonic electrons.}
\label{fig_elec_raa_cent}
\end{figure}

From Run-4, we have collected a significantly larger data sample and have reduced the converter material near the beam-pipe,
thus reducing the radiation length before our tracking detectors from 1.3\% to 0.4\%.  We use two different methods for extracting the non-photonic 
electron distribution from the initially measured inclusive electron sample.  One method is to subtract off all known
photonic contributions, using our own measurements of the $\pi^{0}$ and $\eta$ as input.  The second method is making use
of a special ``converter run'' where we purposely increase the radiation length around the beam pipe.  We can then
compare the inclusive electron yield between this ``converter run'' and normal running to determine the photonic
contribution and subtract it away.  The first method has larger systematics at low $p_{T}$ since the ratio of non-photonic
to photonic electrons is small.  The second method works quite well at low $p_{T}$, but is statistics limited at
high $p_{T}$ due to the short time duration of the ``converter run''.  Thus, the two methods are quite complementary
and also agree quite well at intermediate $p_{T}$ where both maintain good accuracy.  In Figure~\ref{fig_elec_spectra} we
show the PHENIX Preliminary results for non-photonic single electrons as a function of $p_{T}$ for various $Au+Au$
centrality selections.  In addition, the curve is the best fit to our non-photonic electron result from proton-proton
reactions, scaled up by the expected number of binary collisions.  Shown in Figure~\ref{fig_elec_raa_cent} is the
nuclear modification factor ($R_{AA}$) for central $Au+Au$ reactions.  We observe a significant suppression that appears
to increase as a function of $p_{T}$ and is strongest for the most central reactions.  

Calculations assuming only radiative charm quark energy loss are able to describe the data with varying 
degrees of success~\cite{armesto,djordjevic}.
However, we note that one can always arbitrarily increase $dN/dy$(gluon) or similarly the $\hat{q}$ value, but then one 
may observe conflicts with light quark/gluon energy loss results or total entropy limits.  
It has been pointed out~\cite{djordjevic}
that beauty meson semi-leptonic decay may contribute significantly to the non-photonic electrons for $p_{T}>3$ GeV/c
and that the dead-cone effect significantly limits bottom quark energy loss.  However, we note that for bottom quarks,
neglecting collisional energy loss, as opposed to radiative, may not be well justified.  Other calculations include only 
collisional energy loss~\cite{teaney}.  Better
experimental constraints on the charm and beauty total cross sections will also be an important ingredient in understanding
where each contributes and at what level.  PHENIX has a preliminary measurement of the Upsilon in proton-proton reactions
which is a start at gauging beauty production.

\begin{figure}[htb]
\vspace*{+0.0cm}
\insertplot{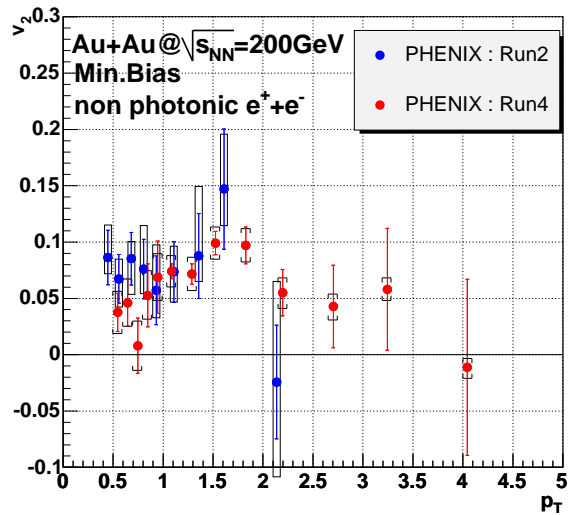}
\vspace*{-0.7cm}
\caption[]{PHENIX Run-2 published and Run-4 preliminary $v_{2}$ for non-photonic electrons as measured in
minimum bias $Au+Au$ 200 GeV reactions.}
\label{fig_elec_flow}
\end{figure}

PHENIX presented preliminary results from Run-4 on the momentum anisotropy ($v_{2}$) for non-photonic electrons, as shown in
Figure~\ref{fig_elec_flow}. 
It is now conclusive that these electrons have a substantial non-zero anisotropy.  
We should note that some care is warranted in interpreting
these results as ``charm flow.''  We believe these electrons are dominated by semi-leptonic decay of charm mesons and
an additional contribution from the decay of beauty mesons at higher $p_{T}$.  Due to the decay kinematics, the electron 
carries only a fraction of the meson $p_{T}$ and the effective $v_{2}$ is reduced for low $p_{T}$ electrons as the $\Delta \phi$ 
between the parent and daughter particle is effectively like an additional reaction plane smearing contribution.  We 
note that the data gives an indication for a decrease in $v_{2}$ above a $p_{T} \approx 2$ GeV/c.  This could be the result of
a decrease in charm quark flow as predicted in ~\cite{teaney} or due to the emergence of beauty contributions.

\subsection{Quarkonia Results}

In studying closed charm or beauty in heavy ion reactions, we are interested in the interaction between
the heavy quark and antiquark and the surrounding medium.  However, we must also keep in mind that although the state
likely begins as partons, it must transform itself into a hadron before being directly observed.  The promise
of insight from quarkonia measurements is tempered by the large number of possible effects impacting their final yield.  
There may be nuclear modification to the incoming parton distributions (e.g. shadowing, anti-shadowing, EMC,...) that
may impact the exact scaling of total heavy flavor production.  After initial creation of the quark antiquark pair, they are
bombarded by the ``back-side'' of the two nuclei.  These nucleons and the quarks and gluons inside them may break up the
heavy $c\overline{c}$ pair - a process referred to as normal nuclear absorption.  Then the $c\overline{c}$ pair can interact
with the surrounding medium either at the partonic or hadronic level.  

The PHENIX experiment has submitted for publication results on $J/\psi$ production in proton-proton and 
deuteron-$Au$ collisions at RHIC
energies over a broad range in rapidity $-2.2 < y < +2.2$~\cite{phenix_jpsi_da}.    In comparing our proton-proton
and deuteron-$Au$ results, we find a very modest suppression (or none within errors) of $J/\psi$ at mid-rapidity 
relative to  binary scaling and a possible larger suppression
(of order 20\%) at forward rapidity giving a hint of gluon shadowing effects.  Future higher statistics deuteron-$Au$
data will be required for more precise conclusions, but the current data are consistent with a $J/\psi$(precursor)-nucleon
breakup cross section of order 1-3 mb.

\begin{figure}[htb]
\vspace*{+0.0cm}
\insertplot{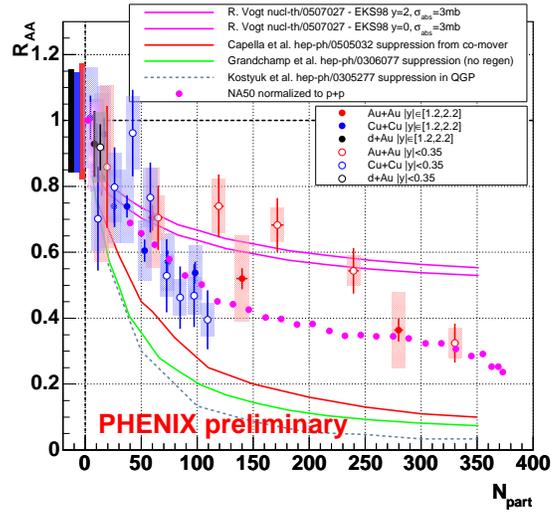}
\vspace*{-0.7cm}
\caption[]{PHENIX Preliminary $Au+Au$ and $Cu+Cu$ 200 GeV nuclear modification $R_{AA}$ for $J/\psi$.
Various theoretical predictions and experimental data as described in the text are shown compared with the PHENIX data.}
\label{fig_jpsi}
\end{figure}

Previously, PHENIX had published only a very low statistics result on $J/\psi$ production in heavy ion
reactions~\cite{phenix_jpsi_aa}.
From the Run-4 and Run-5 high statistics samples, the PHENIX experiment has presented preliminary 
results on $J/\psi$ production in $Au+Au$ and $Cu+Cu$ reactions as a 
function of collision centrality, transverse momentum and rapidity.  All results are shown together in Figure~\ref{fig_jpsi}
in terms of the nuclear modification factor $R_{AA}$ as a function of the number of participating nucleons.  We
observe a suppression of $J/\psi$ yields relative to binary scaling.  The suppression measured at mid-rapidity (via the dielectron channel) is comparable within statistical and systematic errors of that measured at
forward rapidity (via the dimuon channel).  

We overlay the PHENIX data with
three different theoretical and experimental comparisons.  First, shown as the upper two red curves are calculations 
assuming only nuclear modification of parton distribution functions (EKS98) and
normal nuclear absorption with a cross section $\sigma=$ 3 mb, which is at the limit of agreement with our
deuteron-$Au$ data~\cite{vogt}.  These calculations for $y=0$ and $y=2$ appear to underpredict the level of suppression for the more
central $Cu+Cu$ and $Au+Au$ data.  Next we show various calculations assuming further suppression of the $J/\psi$
due to comover absorption or disassociation due to screening~\cite{jpsi_theory}.  These three particular 
calculations were all matched to the $J/\psi$ suppression observed at lower energies, and since the RHIC energy
density is a factor of 2-3 higher, they all substantially overpredict the level of suppression.  Finally, we show the
experimental data from lower energy from experiment NA50, normalized to yield $R_{AA}=1$ for the most
peripheral events~\cite{na50}.  Although one must take seriously the current PHENIX systematic errors, the general agreement
with the lower energy centrality dependence is striking.  
There are a variety of theoretical calculations invoking $J/\psi$ regeneration or coalescence~\cite{regeneration}.
These calculations qualitatively predict very large initial suppression of $J/\psi$, which is compensated
for by later re-formation.  These calculations may give a better description of the data, but must be
checked in the context of the $J/\psi$ proton-proton cross section and the input charm quark cross section
and distributions~\cite{ralf}.

 
\begin{figure}[htb]
\vspace*{+0.0cm}
\insertplot{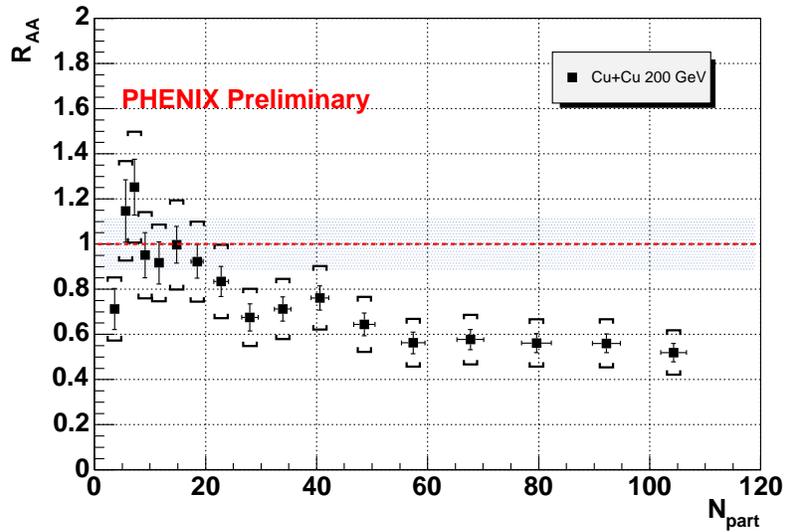}
\vspace*{-0.7cm}
\caption[]{PHENIX Preliminary $R_{AA}$ for $J/\psi$ in $Cu+Cu$ 200 GeV reactions.}
\label{fig_cucu}
\end{figure}

We have also presented PHENIX preliminary transverse momentum and rapidity distributions of $J/\psi$
from $Au+Au$ and $Cu+Cu$ reactions.  We find no large modifications of these distributions in
comparing proton-proton to heavy ion reactions, but further quantification of this conclusion requires
pushing down our current systematic errors.  It is notable that since charm quarks show a suppression at
high $p_{T}$, one might expect $J/\psi$ regeneration to lead to a significant distortion of the 
transverse momentum distribution.  Predictions of narrower rapidity distributions have also been made.
We show in Figure~\ref{fig_cucu} $R_{AA}$ in 17 bins in centrality from our $Cu+Cu$ data set, which emphasizes that we have excellent statistics to explore various dependencies.  For example, it was suggested that perhaps only the $\chi_{c}$ is
suppressed and that is why the NA50 and PHENIX suppression patterns appear similar.  If this were
the case one might naively expect the suppression onset to occur at RHIC energies in mid-central $Cu+Cu$,
whereas our data show a slow onset of the full suppression seen for the most central events.  Whether the $J/\psi$ is really suppressed in medium and then
regenerated or is never much suppressed at all is a question we hope to answer with the reduction of our 
current statistical and systematic errors and future measurements.

\section{Conclusions}\label{concl}
In summary, there is a wealth of new PHENIX data on heavy quarks and heavy quarkonia.  We will work
hard to push these results to submitted publications. Charm is a very optimal probe of thermalization and
properties of the medium, but the price for this may well be the loss of a probe via quarkonia
for deconfinement.
 
\section*{Acknowledgments}
We wish to thank the organizers of this excellent workshop and for the opportunity to have
extended back and forth discussions on many interesting topics.

\vfill\eject
\end{document}